
%
%
\magnification=\magstep 1
\font\bigbf=cmbx12 \font\bigbigbf=cmbx12 scaled \magstep1\font\eightrm=cmr8
\font\eightit=cmti8 \font\refs=cmcsc10
\font\mib=cmmib10\font\mibsev=cmmib10 scaled 700\font\mibfiv=cmmib10 scaled 500
\font\mbsy=cmbsy10\font\mbsysev=cmbsy10 scaled 700\font\mbsyfiv=cmbsy10 scaled
500\font\bol=cmbx10\font\bolsev=cmbx7\font\bolfiv=cmbx5
\def\bold{
   \textfont0=\bol \scriptfont0=\bolsev \scriptscriptfont0=\bolfiv
   \textfont1=\mib \scriptfont1=\mibsev \scriptscriptfont1=\mibfiv
   \textfont2=\mbsy \scriptfont2=\mbsysev \scriptscriptfont2=\mbsyfiv
   \textfont3=\tenex \scriptfont3=\tenex \scriptscriptfont3=\tenex}
\def\today{\number\day\ \ifcase\month\or January\or February\or March\or April
\or May\or June\or July\or August\or September\or October\or November\or
December\fi\space \number\year}
\headline={\ifodd\pageno\lefthead\else\righthead\fi}
\def\righthead{\hfil{\eightit RSB in the Random Replicant Model}\hfil}
\def\lefthead{\ifnum\pageno=45\firsthead\else\otherhead\fi}\def\firsthead{\hfil}
\def\otherhead{\hfil{\eightrm P Biscari \& G Parisi}\hfil}
\footline={\ifnum\pageno>44\firstfoot\else\foot\fi}
\def\firstfoot{\hfil}\def\foot{\hfil{\rm\folio}\hfil}
\catcode`\@=11\def\@ifundefined#1{\expandafter\ifx\csname#1\endcsname\relax}
\def\rfeq#1{\@ifundefined{eq@#1}\message{Eq. #1 is undefined - }
\else\csname eq@#1\endcsname\fi}
\def\eq#1{\avantin\@namedef{eq@#1}{\numcap.\number\nqe}
\immediate\write\aux{
\string\@namedef\string{eq@\noexpand#1\string}\string{\numcap.\number
\nqe\string}}\eqno{\hbox{(\rfeq{#1})}}}
\def\noeq#1{\avantin\@namedef{eq@#1}{\numcap.\number\nqe}
\immediate\write\aux{
\string\@namedef\string{eq@\noexpand#1\string}\string{\numcap.\number
\nqe\string}}}
\def\rfbib#1{\@ifundefined{bb@#1}\message{Ref. #1 is undefined - }
\else\csname bb@#1\endcsname\fi}
\def\bib#1{\avantib\@namedef{bb@#1}{\number\nbib}
\immediate\write\aux{
\string\@namedef\string{bb@\noexpand#1\string}\string{\number\nbib\string}}
\ited{[\rfbib{#1}]}}
\def\@namedef#1{\expandafter\xdef\csname #1\endcsname}
\def\sect#1{\bigbreak\bigbreak\noindent\avansez{\bigbf
\numcap.~#1}\bigskip\nqe=0}
\newread\aux
\openin\aux=aux.tex
\ifeof\aux\message{File AUX not found}\else\input aux\fi
\closein\aux\immediate\openout\aux=aux.tex
\newcount\nbib\newcount\nqe\newcount\nsez
\def\avantin{\global\advance\nqe by 1}
\def\avansez{\global\advance\nsez by 1}
\def\avantib{\global\advance\nbib by 1}
\catcode`\@=12\nsez=0\nqe=0\nbib=0
\def\rea{{{\rm l} \kern -.15em {\rm R} }}\def\hamjx{{\cal H}_J [\xb]}
\def\mcr{{\mu_{{\rm cr}}}}\def\O{{\cal O}}\def\M{{\cal M}}
\def\Int{\mathop{\rm Int}\nolimits}
\def\J{$\{ J_{ij} \}$}\def\lp{\left}\def\rp{\right}\def\dps{\displaystyle}
\def\d{{\rm d}}\def\e{{\rm e}}\def\Trn{\mathop{\rm Tr}_n\nolimits}
\def\ref#1#2#3#4#5{{\refs #1}; {\it #2} {\bf #3} (#4), #5.}
\def\hand{\hangindent=0.7cm}
\def\textind#1{\llap{#1\enspace}\ignorespaces}\def\aac{\aa_{{\rm c}}}
\def\ited{\smallskip\noindent\hskip 0.7cm\hand\textind}
\def\zz{\tilde z}\def\ttt{\tilde t}\def\aa{\tilde a}\def\lala{\tilde\lambda}
\def\medz#1{\langle #1 \rangle_z}\def\Lrs{{\cal L}_{{\rm RS}}}
\def\rL{{\rm L}}\def\frs{f_{{\rm RS}}}
\def\fh{f_{{\rm H}}}\def\ovu#1{\overline{\medz{#1}}}
\def\ovd#1#2{\overline{\medz{#1}\medz{#2}}}\def\ovq#1{\overline{\medz{#1}^2}}
\def\ovc#1{\overline{\medz{#1}^3}}\def\ovdq#1#2{\overline{\medz{#1}^2\medz{#2}}}

\def\at{\tilde{\alpha}}
\def\gt{\tilde{\gamma}}
\def\zr{z_r}
\def\Lh{{\cal L}_{{\rm H}}} \def\peta{ {\rm
P}(z,\zr)}\def\medzz#1{\left\langle #1 \right\rangle_{(z,z_r)}}
\def\medr#1{\left[ #1 \right]_{z}}\def\fd{f^{(2)}}\def\ft{f^{(3)}}

\def\rr{{\tilde r}}
\def\derq{{\delta\over\delta Q(x)}}\def\dxib{\hbox{$\delta\xi\bold$}}
\def\Qb{\hbox{$Q\bold$}}\def\lab{\hbox{$\lambda\bold$}}\def\xb{\hbox{$x\bold$}}
\def\numcap{\number\nsez}
\def\sxb{\hbox{$\scriptstyle{x}\bold$}}\def\sQb{\hbox{$\scriptstyle{Q}\bold$}}
\def\slab{\hbox{$\scriptstyle{\lambda}\bold$}}
%
%
\pageno=45
\null\vfil
{\baselineskip=24truept \centerline{\bigbigbf Replica Symmetry Breaking}
\centerline{\bigbigbf in the Random Replicant Model}}
\vfil\noindent
{\refs P Biscari${}^{(1)}$ \& G Parisi${}^{(2)}$}
\bigskip
\item{${}^{(1)}$} Dip.~Matematica, Universit\`a di Pisa, Via Buonarroti 2,
56127 Pisa, Italy\hfil\break
biscari@vaxsns.sns.it
\item{${}^{(2)}$} Dip.~Fisica, Universit\`a di Roma ``La Sapienza'',
P.le A.~Moro, 00187 Roma, Italy\hfil\break
parisi@vxrm70.roma1.infn.it
\vfil\noindent
{\bf Abstract. } We study the statistical mechanics of a model
describing the coevolution of species interacting in a random way. We
find that at high competition replica symmetry is broken. We solve the
model in the approximation of one step replica symmetry breaking and
we compare our findings with accurate numerical simulations.
\vfil\noindent
Short title: {\it RSB in the Random Replicant Model}
\bigskip\noindent
Submitted to: {\it Journal of Physics A: Mathematical and General}
\bigskip\noindent
Date: \today
\eject
\pageno=1
\sect{Introduction}

The replicant models study the coevolution of sets of interacting
species able to reproduce themselves: they have a huge number of
applications in biologic and optimization problems [\rfbib{scsi} --
\rfbib{mgk}]. In this paper we study a non-deterministic evolution: we
consider a system of replicants which evolve with random interactions.

The model, introduced by S.~Diederich and M.~Opper in [\rfbib{diop}],
is defined as follows. Given $N$ species, let $x_i/N$ be the
concentration of the $i$-th family in the system. The real variables
$\{x_i\in\rea,$ $i=1,\dots,N\}$ are then subject to the constraints
$$\sum_{i=1}^N x_i = N,\qquad\qquad x_i\ge 0\quad\forall i=1,\dots,N.
\eq{norm}$$

The interactions between different species are described through
a fitness functional $F_J(\xb)$ that must be maximized at equilibrium.
Typically, $F_J$ is chosen as a quadratic function of the
concentrations, that is equivalent to take into account only pair
interactions between the species:
$$F_J(\xb)=-\hamjx = -\sum_{i<j=1}^N J_{ij}\;x_i\;x_j - a\sum_{i=1}^N
x_i^2,\eq{def}$$
where the parameters \J\ are chosen at random from the Gaussian
probability distribution
$$P(J_{ij}) = \sqrt{{N\over\pi J^2}} \exp\lp(-{NJ_{ij}^2\over J^2}\rp)
\eq{pdij}$$
like in the Sherrington-Kirkpatrick model of spin glasses [\rfbib{edan},
\rfbib{shki}]. The control parameter $a$ has the aim of limiting the growth
and the supremacy of one single species: for big values of $a$,
the growth of all the species is
strongly limited by the factor $ax_i^2$; in that case, the random
interactions become negligible and the equilibrium configuration is
$$x_i^{{\rm eq}} \simeq 1 \qquad \forall i=1,\dots,N,\qquad \lp(
a\gg J\rp)\eq{biga}$$
almost independently from the interactions between the species.
Instead, for small values of $a$, the pair interactions play a central
r\^ole and a few species prevail among the others.
Analytically, this model differs from the SK spin glass in that we
impose the constraint (\rfeq{norm}): the spins are then allowed to
take any real value, but the total magnetization is fixed.

In section 2 we show how it is possible to solve the random replicant
model within the replica formalism. In sections 3 and 4, we analyze
the replica symmetric solution and its stability and in section 5 we
perform the first step of the hierarchical replica symmetry breaking.
The biological applications of the results are found in the limit
$T\to 0^+$ because the fitness functional $F_J$ introduced in the last
section is, a minus sign apart, the low temperature limit of the free
energy.

The study of the stability of the replica symmetric solution will
show that, at zero temperature, the replicant model exhibits a phase
transition to a glassy phase when $a$ crosses a certain value $a_c$.
The replica symmetry breaking which occurs in the glassy phase
$(a<a_c)$ implies the breakdown of the ergodicity of the system: when
$a$ becomes small, the evolution of the system depends strongly on the
initial conditions, and in general we will not be able to make any
precise prediction on the equilibrium state of the system.

{}From the biological point of view, the glassy phase is the unstable
one: in the high $a$ phase, a single equilibrium state exists, and the
system is able to recover its equilibrium configuration after any
external change of the concentrations of its elements;
on the contrary, in the glassy phase, the same perturbation can change
drastically the final configuration of the system, if it is led to a
different ergodic region of the phase space. Here however we study
only the properties of the statics associated to Hamiltonian
(\rfeq{def}) and we do not consider the dynamics of a system leading
to this equilibrium distribution.

\sect{The Random Replicant Model: analytical solution}

Now we derive the expression for the quenched free energy density of
the random replicant model. In this and the next section, we follow
closely [\rfbib{diop}].
The evolution of the system is ruled by the Hamiltonian (\rfeq{def});
averaging over all the possible choices of the \J, the quenched free
energy of the system is given by
$$ -\beta N f = \int \prod_{i<j} \d J_{ij}\; P(J_{ij})\; \ln
\sum_{[\sxb]} \exp\lp( - \beta \hamjx \right).\eq{fqu}$$
To compute (\rfeq{fqu}), we use the replica method [\rfbib{shki} --
\rfbib{parc}] introducing a set $\{\lambda_\alpha,$
$\alpha=1,\dots,n\}$ of Lagrange multipliers which ensure the
normalization condition (\rfeq{norm}) in each of the $n$
replicas. With standard calculations [\rfbib{bisc}], we arrive at the
following expression for $f$:
$$-\beta f = \lim_{n\to 0^+}\max_{\sQb,\slab} \lp[-{1\over n}\sum_{\alpha
\gamma} Q_{\alpha\gamma}^2+{1\over n}\sum_\alpha\lambda_\alpha+{1\over n}\ln
\Trn\exp \rL(\Qb,\lab,\xb)\rp],\eq{ffin}$$
where
$$\rL(\Qb,\lab,\xb):= -\beta a \sum_\alpha x_\alpha^2 - \sum_\alpha
\lambda_\alpha x_\alpha+\beta J \sum_{\alpha\gamma}
Q_{\alpha\gamma}x_\alpha x_\gamma\; ;\eq{dfl}$$
\Qb\ and \lab\ are respectively an $n\times n$ matrix and an
$n$-dimensional vector; $\{x_\alpha,$ $\alpha=1,\dots,n\}$ is a new set of
real positive variables, and $\Trn$ denotes the integral over all possible
values of the $x_\alpha$'s.

{}From (\rfeq{ffin}) and (\rfeq{dfl}) we find that the stationarity equations
for $f$ are:
$$\cases{\dps{\eqalign{Q_{\alpha\gamma} &= {\beta J\over 2}\;
{\Trn x_\alpha x_\gamma \exp \rL(\Qb,\lab,\xb) \over \Trn \exp
\rL(\Qb,\lab,\xb)}\qquad
 \forall \alpha,\gamma=1,\dots,n;\cr
1&= {\Trn x_\alpha \exp \rL(\Qb,\lab,\xb) \over \Trn \exp \rL(\Qb,\lab,\xb)}
\qquad\qquad\qquad\forall\alpha=1,\dots,n.\cr}}&\cr}\eq{sols}$$
The remaining sections are devoted to the study of the solutions of
these stationarity conditions.

\sect{Replica symmetric solution}

Both the free energy density and the stationarity equations above are
invariant under the action of the group ${\cal S}_n$ of permutations between
the $n$ replicas. This implies that at least one of the solutions of
(\rfeq{sols}) is invariant under ${\cal S}_n$, so that the first ansatz that
is to be tried is certainly the symmetric one, which is given by:
$$Q_{\alpha\gamma} = q \;\delta_{\alpha\gamma}+t\qquad{\rm and}\qquad
\lambda_\alpha = \lambda.\eq{simso}$$
Introducing (\rfeq{simso}) into (\rfeq{ffin}), and denoting by $\frs$ the
resulting the free energy density we have, after the manipulations
described in [\rfbib{bisc}],
$$-\beta \frs = \max_{q,\ttt,\lala} \lp[q^2+2\beta J\; q\ttt-\beta J\;\lala-
\overline{\ln\lp(\int_0^{+\infty}\d x\exp\Lrs(q,\ttt,\lala,x,z)\rp)}\ \rp],
\eq{frf}$$
where
$$\Lrs(q,\ttt,\lala,x,z):= -\beta J \lp[(\aa-q) x^2 -\big(2z\sqrt{ \ttt}-
\lala\big)x\rp],\eq{frlf}$$
$\ttt:=t/(\beta J)$, $\lala:=\lambda/(\beta J)$, $\aa:=a/J$, and we have
introduced the notation
$$\overline {G(z)} := {1\over \sqrt{\pi}} \int_{-\infty}^{+\infty} \d z\;
\e^{-z^2}\; G(z).\eq{over}$$
In a similar way, the stationarity equations become:
$$\cases{\dps{\ovu{x} = 1,}&\cr\cr
\dps{\ovq{x} = 2 \ttt, } & \cr\cr
\dps{\overline{\medz{x^2}-\medz{x}^2}= {2 q \over \beta J },}&\cr}\eq{solsr}$$
where
$$\medz{\;G(x)\;} := {\dps{\int_0^{+\infty} \d x\; G(x)\; \exp
\Lrs(q,\ttt,\lala,x,z)} \over \dps{\int_0^{+\infty} \d x\; \exp
\Lrs(q,\ttt,\lala,x,z)} }. \eq{ntz}$$

The low temperature limit of the symmetric solution, first studied by
Diederich and Opper [\rfbib{diop}], is particularly interesting because
it allows us to prove analytically the existence of a second order
transition to a glassy phase, as we will show in the next section.
Introducing the parameter
$$\zz:= {\lala\over 2\sqrt{\ttt}}\; ,\eq{zzz}$$
the stationarity equations (\rfeq{solsr}) become:
$$\cases{\dps{4\;q\;(\aa-q)= {1\over\sqrt{\pi}}\int_{\zz}^{+\infty} \d z\;
\e^{-z^2}\;;}&\cr\cr
\dps{ 4 \;\ttt\;(2q-\aa)=\lala\;;}&\cr\cr
\dps{{\sqrt{\ttt}\;\e^{-\zz^2}\over \sqrt{\pi}}-2(\aa-q)={\lala\over\sqrt{\pi}}
\int_{\zz}^{+\infty} \d z\; \e^{-z^2}},&\cr}\eq{lw3}$$
leading to $\frs = 2J\ttt(\aa-2q)$.
Figure 1 shows how $q$, $\ttt$, $\lala$, and $\frs$ behave as functions of
$\aa$ in this limit. We give also the approximate expressions of these
parameters in two particularly interesting cases: the ``classical'' regime
$(\aa\gg 1)$, and the neighbourhood of the critical point $\aac=1/\sqrt{2}$.

\midinsert\noindent
\null\vskip 14truecm\noindent
Figure 1. Numerical solutions of the replica symmetric equations in the low
temperature limit
\smallskip
\endinsert

In the former region the equilibrium configurations
become trivial, with $x_i^{{\rm eq}}\simeq 1$, $\forall i$. The replica
symmetric solution, which we will prove to be stable in this region,
predicts:
$$\eqalign{q &= {1\over 2}\lp(\aa-\sqrt{\aa^2-1}\rp)+\O\big(\e^{-\aa^2}\big),
\cr
\ttt &= {1\over 4}\lp(1+{\aa\over\sqrt{\aa^2 -1}}\rp)+\O\big(\e^{-\aa^2}\big),
\cr
\lala &= -\aa - \sqrt{\aa^2-1}+\O\big( \e^{-\aa^2}\big),\cr}\eq{lw65}$$
and the free energy density becomes
$$\frs^o ={1\over 2}\lp(\aa+\sqrt{\aa^2-1}\rp)+\O\big( \e^{-\aa^2}\big).
\eq{lw7}$$
Instead, the latter is the transition point to the glassy phase, as we
will show below; in its neighbourhood we have:
$$\eqalign{q &= {\sqrt{2}\over 4} - {\sqrt{2}\over 2}(\pi-2)(\aa-\aac)^2+
o\big( (\aa-\aac)^2\big),\cr
\ttt &= {\pi\over 2} - \sqrt{2} \pi (\pi-2)(\aa-\aac)+\pi^2(3\pi-8)
(\aa-\aac)^2+o\big( (\aa-\aac)^2\big),\cr
\lala &= -2\pi(\aa-\aac)+2\sqrt{2}\pi(\pi-2)(\aa-\aac)^2+
o\big((\aa-\aac)^2\big),\cr
\frs^o &= \pi(\aa-\aac)-\sqrt{2}\pi(\pi-2)(\aa-\aac)^2+
o\big( (\aa-\aac)^2\big).\cr}\eq{lw6}$$

Finally, figure 2 shows the numerical results that we obtained for the
order parameters $q$ and $t$ by solving equations (\rfeq{solsr}) for
different finite values of $\beta$.

\midinsert\noindent
\null\vskip 9truecm\noindent
Figure 2. Numerical solutions of the replica symmetric equations at finite
temperature
\smallskip
\endinsert

\sect{Instability of the symmetric solution}

In the preceding section we have shown that equations (\rfeq{sols})
admit a symmetric solution, but we must also check the Hessian of the
free energy to determine whether our solution is a minimum of $f$
or just a saddle-point. To find the eigenvalues of the Hessian we
generalize the calculus made by De Almeida and Thouless [\rfbib{alth}]
for the SK model of spin glass: let
$$\eqalign{Q_{\alpha\gamma}&=(q+\delta q_\alpha)\;\delta_{\alpha\gamma} + t+
\delta t_{\alpha\gamma},\qquad {\rm with}\ \cases{\delta t_{\alpha\alpha}=0,&
\cr\delta t_{\alpha\gamma}=\delta t_{\gamma\alpha};&\cr}\cr
\lambda_{\alpha}&=\lambda+\delta\lambda_\alpha.\cr}\eq{f1}$$

If we denote by $\dxib$ the vector ($\delta\lambda\bold$; $\delta q\bold$;
$\delta t\bold$) and we substitute (\rfeq{f1}) in (\rfeq{ffin}) we
obtain, after some tedious calculations [\rfbib{bisc}], that the
second order term in the expansion of $f$ in terms of $\dxib$ is given
by $-\beta\delta_2 f ={1\over 2}\; \dxib^T \cdot \M \;\dxib$,
where $\M$ is a real symmetric matrix with the following fourteen
different types of elements:
$$\eqalign{ A:=\M_{\delta\lambda_\alpha\delta\lambda_\alpha}&=
\lp(\ovu{x^2}-\ovu{x}^2\rp),\cr
B:=\M_{\delta\lambda_\alpha\delta\lambda_\gamma}&=\lp(\ovq{x}-
\ovu{x}^2\rp),\cr
C:=\M_{\delta\lambda_\alpha\delta q_\alpha}&=-\beta J\lp(\ovu{x^3}-
\ovu{x^2}\;\ovu{x}\rp), \cr
D:=\M_{\delta\lambda_\alpha\delta q_\gamma}&=-\beta J\lp(\ovd{x^2}{x}-
\ovu{x^2}\;\ovu{x}\rp),\cr
E:=\M_{\delta q_\alpha\delta q_\alpha}&=-2+\beta^2 J^2\lp(\ovu{x^4}-
\ovu{x^2}^2\rp),\cr
F:=\M_{\delta q_\alpha\delta q_\gamma}&=-\beta^2 J^2\lp(\ovq{x^2}-
\ovu{x^2}^2\rp),\cr
G:=\M_{\delta\lambda_\alpha\delta t_{\alpha\gamma}}&=-\beta J
\lp(\ovd{x^2}{x}- \ovq{x}\;\ovu{x}\rp),\cr
H:=\M_{\delta\lambda_\alpha\delta t_{\gamma\delta}}&=-\beta J
\lp(\ovc{x}- \ovq{x}\;\ovu{x}\rp),\cr
I:=\M_{\delta q_\alpha\delta t_{\alpha\gamma}}&=\beta^2 J^2
\lp(\ovd{x^3}{x}- \ovu{x^2}\;\ovq{x}\rp),\cr
J:=\M_{\delta q_\alpha\delta t_{\gamma\delta}}&=\beta^2 J^2
\lp( \overline{\medz{x^2}\medz{x}^2}-\ovu{x^2}\;\ovq{x}\rp),\cr
K:=\M_{\delta t_{\alpha\gamma}\delta t_{\alpha\gamma}}&=-2+\beta^2 J^2
\lp( \ovq{x^2} - \ovq{x}^2\rp),\cr
K':=\M_{\delta t_{\alpha\gamma}\delta t_{\gamma\alpha}}&=\beta^2 J^2
\lp( \ovq{x^2} - \ovq{x}^2\rp)=K+2\; ,\cr
L:=\M_{\delta t_{\alpha\gamma}\delta t_{\gamma\delta}}&=\beta^2 J^2
\lp( \overline{\medz{x^2}\medz{x}^2} - \ovq{x}^2\rp),\cr
M:=\M_{\delta t_{\alpha\gamma}\delta t_{\delta\eta}}&=\beta^2 J^2
\lp( \overline{\medz{x}^4} - \ovq{x}^2\rp).\cr}\eq{f11}$$
Furthermore, $\M$ has three different types of eigenvectors:
\ited{(i)} symmetric eigenvectors of the type
$$\dxib=(\ell,\dots,\ell;\rho,\dots,\rho;
\tau,\dots,\tau);\eq{f13}$$
\ited{(ii)} 1-asymmetry eigenvectors, with:
$$\eqalign{\delta\lambda_\alpha &= \cases{\ell_1 & if $\alpha=\at$ \cr
\ell_0 & otherwise,} \cr
\delta q_\alpha &= \cases{\rho_1 & if $\alpha=\at$ \cr
\rho_0 & otherwise,} \cr
\delta t_{\alpha\gamma} &= \cases{\tau_1 & if $\alpha=\at$ or $\gamma=\at$\cr
\tau_0 & otherwise;} \cr} \eq{f16}$$
\ited{(iii)} 2-asymmetries eigenvectors, of the type:
$$\eqalign{\delta\lambda_\alpha &= \cases{\ell_1 & if $\alpha=\at$ or
$\alpha=\gt$\cr \ell_0 & otherwise, \cr} \cr
\delta q_\alpha &= \cases{\rho_1 & if $\alpha=\at$ or $\alpha=\gt$\cr
\rho_0 & otherwise,\cr} \cr
\delta t_{\alpha\gamma} &= \cases{\tau_2 & if $\alpha\gamma=\at\gt$ or
$\alpha\gamma=\gt\at$\cr
\tau_0 & if $\alpha\ne\at$, $\alpha\ne\gt$, $\gamma\ne\at$ and
$\gamma\ne\gt$ \cr
\tau_1 & otherwise.} \cr} \eq{f19}$$

The eigenvalues of $\M$ must be negative in order to ensure the stability
of the symmetric ansatz. The biggest among the eigenvalues associated with
the families above comes from the 2-asimmetries eigenvectors, and is given
by [\rfbib{bisc}]
$$\mcr={1\over 2}\big(K+K'-2L+M\big)=-1+\beta^2\overline{\lp(\medz{x^2}-
\medz{x}^2\rp)^2}.\eq{f22}$$
In the low temperature limit $\mcr$ can be easily computed, and is equal
to
$$\mcr={2q-\aa\over \aa-q}.\eq{q1}$$
In particular, as figure 3 shows, $\mcr$ becomes positive when $\aa<\aac$:
$$\mcr=\pi(\aa-\aac)-\sqrt{2}\pi(\pi-2)(\aa-\aac)^2+
o\big((\aa-\aac)^2\big).\eq{y3}$$

\midinsert\noindent
\null\vskip 9truecm\noindent
Figure 3. Critical eigenvalue at zero temperature. The symmetric solution
becomes unstable when it is positive.
\smallskip
\endinsert

\sect{Replica symmetry breaking}

Having proved the instability of the symmetric solution, we must now search
a more general ansatz to describe the system when $\aa<\aac$. To obtain it,
we will follow the guidelines of the hierarchical ansatz of spin glasses
[\rfbib{para}, \rfbib{parb}, \rfbib{parc}]. In this paper we study only the
first step of the replica symmetry breaking, testing order parameter
matrices of the type
$$Q_{\alpha\gamma} =\cases{ t & if $\dps{\Int\lp({\alpha\over \eta}\rp)\ne
\Int\lp({\gamma\over \eta}\rp)}$,\cr\cr
t+r & if $\dps{\Int\lp({\alpha\over \eta}\rp)= \Int\lp({\gamma\over \eta}\rp)}$
but $\alpha\ne\gamma$, \cr\cr
q+t+r & if $\alpha=\gamma$. \cr}\eq{y10}$$
This ansatz can be improved by iterating the breaking scheme in
all the blocks introduced in the first step, but we will see that even
a single breaking improves drastically the symmetric predictions. We recall
that, in the limit $n\to 0^+$, the hierarchical parametrization can be
written in terms of an order parameter function $Q(x)$, defined in the
interval $x\in [0,1]$, which, at this point of symmetry breaking, is equal
to
$$Q(x)=\cases{t & if $x\in[0,\eta)$, \cr t+r & if $x\in (\eta,1)$. \cr}
\eq{func}$$
In (\rfeq{func}) we have omitted the diagonal term containing $q$
(corresponding to $Q(1)$), because it involves the term in the
Hamiltonian that contains $a$ and it can always be treated separately.
The introduction of the breaking parameters $\eta$ and $r$ changes the free
energy density as follows:
$$\eqalign{-\beta \fh &= \max_{q,t,r,\lambda,\eta}
\Big[-(q+t+r)^2 - (\eta-1)(t+r)^2+\eta t^2+\lambda+\cr
&+{1\over \eta}\int_{-\infty}^{+\infty}\!\!\d z{\e^{-z^2}\over\sqrt{\pi}}
\ln\int_{-\infty}^{+\infty}\!\!\d\zr{\e^{-\zr^2}\over\sqrt{\pi}}\lp(
\int_0^{+\infty}\!\!\d x\exp\Lh(q,t,r,\lambda,x,z,\zr)\rp)^\eta\Big],\cr}
\eq{y14}$$
with
$$\Lh(q,t,r,\lambda,x,z,\zr):= -\beta J (\aa-q) x^2 +
\big(2z\sqrt{\beta Jt}+2\zr\sqrt{\beta Jr}-\lambda\big) x.\eq{y15}$$

The stationarity equations related to $\fh$ become:
$$\eqalign{&1 = \overline{ \medr{\medzz{x}}}\; ,\cr
&t = {\beta J\over 2}\overline{\medr{\medzz{x}}^2}\; ,\cr
&r = {\beta J\over 2}\overline{\Big( \medr{\medzz{x}^2 } - \medr { \medzz{x} }
^2 \Big)}\; , \cr
&q= {\beta J\over 2}\overline{\Big( \medr{\medzz{x^2}} - \medr { \medzz{x}^2
} \Big)}\; , \cr
&\eta^2 r\; (r+2t) = \overline{\lp( \eta \medr{ \ln \int_0^{+\infty}\!\!\!\!
\d x\;\Lh}  - \ln \int_{-\infty}^{+\infty}\!\!\!\!\d \zr
{e^{-\zr^2}\over\sqrt{\pi}} \peta \rp)}\; ,\cr} \eq{y19}$$
where
$$\eqalign{\peta&:=\lp(\int_0^{+\infty}\d x\; \Lh(q,t,r,\lambda,x,z,
\zr)\rp)^\eta, \cr
\medzz{\;\cdot\;} &:= {\dps{\int_0^{+\infty} \d x\; \cdot\;
 \Lh(q,t,r,\lambda,x,z,\zr)} \over  \dps{\int_0^{+\infty}\d x \;
\Lh(q,t,r,\lambda,X,z,\zr) } }\;,\cr
\medr{\;\cdot\;} &:= {\dps{\int_{-\infty}^{+\infty} \d \zr \;
{e^{-\zr^2}\over\sqrt{\pi}} \;\; \cdot\;\;\peta} \over \dps{\int_{-\infty}
^{+\infty}\d \zr\; {e^{-\zr^2}\over\sqrt{\pi}} \;\peta} }. } \eq{y18}$$
Solving numerically equations (\rfeq{y19}) in the low temperature limit, we
find that the product $\eta r$ of the two breaking parameters remains
finite in the $\beta\to +\infty$ limit, and that it becomes different
from zero as soon as $\aa<\aac$, as it is shown in figure 4.

\midinsert\noindent
\null\vskip 9truecm\noindent
Figure 4. Numerical solutions obtained for the product of the breaking
parameters $\eta$ and $r$ at zero temperature.
\smallskip
\endinsert

In figure 5 we show one of the results that we found in the
numerical simulations that we have performed on this model, and that we
will describe in more detail elsewhere: the triangles
represent the free energy density obtained from the simulations at
zero temperature, the continuous line corresponds to the replica symmetric
prediction, and the dashed line illustrates the broken symmetry results.
Figure 5 clearly shows how the first step of the replica symmetry breaking
improves the symmetric predictions, even if it fails when $\aa$ goes to
zero.

\midinsert\noindent
\null\vskip 9truecm\noindent
Figure 5. Improvement led by the 1-step replica symmetry broken solution in
the prediction for the free energy at zero temperature.
\smallskip
\endinsert

To conclude the study of the replica symmetry broken solution we will now
show that, in the low temperature limit, $\eta$ goes to zero as
$\O\big(\beta^{-1}\big)$, so that the breaking parameter $r$ scales
as $t$ and $\lambda$ do, {\it i.e.} that $r=\beta \rr$ with $\rr$
finite as $T$ goes to zero.
We will prove this result near the critical value of $a$ where we
have a better analytic control. To this end, we push our expansion
of $f$ to the third order in $\delta\lambda\bold$, $\delta q\bold$,
$\delta t\bold$, obtaining
$$-\beta f=-\beta \frs+\lim_{n\to 0^+}{\fd+\ft\over n}.\eq{st4}$$
The second order term $\fd$ was studied in the preceding section;
considering the third order term as a functional of the order
parameter function $Q(x)$, and neglecting higher order terms in $n$,
the stationarity equation
$$\derq f[q] =0, \eq{se2}$$
that must be verified $\forall x\in[0,1]$, can be given the form
[\rfbib{bisc}]:
$$\eqalign{ 2 \dot Q(x)\Big\{&1-\beta^2 \big(\ovu{x^2}-\ovq{x}\big)^2+\cr
&+2\ell\beta^2 \big(\ovu{x^3}-3\ovd{x^2}{x}+2\ovc{x}\big)\big(\ovu{x^2}-
\ovq{x}\big) + \cr
&-2\rho \beta^3 \big(\ovu{x^4}-2\ovd{x^3}{x}-\ovq{x^2}+2\ovdq{x}{x^2}\big)
\big( \ovu{x^2}- \ovq{x}\big) +\cr
&+2\beta^3 Q(x)\bigg(  2\big(\ovu{x^2}- \ovq{x}\big)^3 x -\big(\ovu{x^3}
-\ovd{x^2}{x}\big)\big(\ovu{x^3}-5\ovd{x^2}{x}+ \cr
&\qquad\qquad+ 4\ovc{x}\big)+4\ovq{x}\big(\ovu{x^2}- \ovq{x}\big)^2\bigg)+\cr
&+ 4\beta^3 \int_0^1 \d x'\;Q(x')\big(\ovu{x^3}-3\ovd{x^2}{x}+
2\ovc{x}\big)\big(\ovu{x^2}- \ovq{x}\big)  +\cr
&+4\beta^3 \int_x^1 \d x'\;Q(x')\big(\ovu{x^2}- \ovq{x}\big)^3\Big\}=0,\cr}
\eq{se3}$$
where
$$\dot Q(x) := {\d Q\over \d x}.\eq{se4}$$

This expression can be derived again with respect to $x$ to obtain a
necessary condition for the equilibrium:
$$\cases{\dot Q(x) = 0 \qquad{\rm or}&\cr
-4\beta^3\bigg[2 x \bigg(\ovu{x^2}- \ovq{x}\bigg)^3 - \bigg(\ovu{x^3}-
3\ovd{x^2}{x}+2\ovc{x}\bigg)^2\bigg]=0.\cr}\eq{se5}$$

This is exactly what we were looking for: the order parameter function
$Q(x)$ must be a constant in all $x\in [0,1]$, except for
$$x=\eta:= {\bigg(\ovu{x^3}-3\ovd{x^2}{x}+2\ovc{x}\bigg)^2 \over
2 \bigg(\ovu{x^2}- \ovq{x}\bigg)^3 },\eq{se6}$$
where a jump can happen. Note that in Ising spin glasses with two spins
interaction no solution of this type can be found, while a similar
phenomenon happens in the Potts model with $p$ components, when $p>4$.

The integrals in (\rfeq{se6}) can be computed in the low temperature limit,
leading to [\rfbib{bisc}]:
$$\eta={ \dps{{1\over \beta^3 J^3(\aa-q)^3} \sqrt{{\aa-q\over\ttt\pi}}
{\e^{-\zz^2} {\rm c}_{{\rm N}}\over\beta J} } \over  \dps {{4q(\aa-q)\over
4\beta^3 J^3(\aa-q)^3}} }={\e^{-\zz^2}
{\rm c}_{{\rm N}}\over\beta J q \sqrt{\ttt\pi(\aa-q)}}, \eq{sw2}$$
where $\zz$ is defined in (\rfeq{zzz}), and the numerical constant
${\rm c}_{{\rm N}}$ can be easily evaluated, and is equal to
0.0167066...
Equation (\rfeq{sw2}) shows that the scaling behaviuor of $\eta$ is
precisely $\eta=\O\big(\beta^{-1}\big)$ when $\beta\to +\infty$.

\sect{Conclusions}

We have shown that in the replicant model replica symmetry is
broken. The predictions based on one step replica symmetry
breaking are in better agrement with the numerical data than
those coming from exact symmetry. Contrary to what happens in
most of the cases also the one step replica symmetry breaking is
not able to capture the behaviour of the system in the limit of
very small $a$. It would be rather interesting to obtain the
results from full replica symmetry breaking in this region. This
task should not be impossible using the techniques of [\rfbib{bisc}].

\bigbreak
{\bigbf References}
\medskip
\bib{scsi}\ref{P Schuster \& K Sigmund}{J.~Theor.~Biol.}{100}{1983}{533}
\bib{hosi}{\refs J Hofbauer \& K Sigmund}; in {\it Evolutionstheorie und
Dynamische Systeme,} ed.\ Parey, (1984).
\bib{peme}{\refs M Peschel \& W Mende}; in {\it The Predator-Prey
Model,} ed. Springer-Verlag,\break (1986).
\bib{mgk}\ref{H M\"uhlenbein, M Gorges-Schleuter \& O Kr\"amer}{Parallel
Computing}{7}{1988}{65}
\bib{diop}\ref{S Diederich \& M Opper}{Phys.~Rev. A}{39}{1989}{4333}
\bib{edan}\ref{S F Edwards \& P W Anderson}{J.~Phys.~F}{5}{1975}{965}
\bib{shki}\ref{D Sherrington \& S Kirkpatrick}{Phys.~Rev.~Lett.}{35}{1975}
{1792}
\bib{kish}\ref{S Kirkpatrick \& D Sherrington}{Phys.~Rev.~B}{17}{1978}{4384}
\bib{alth}\ref{J R L de Almeida \& D J Thouless}{J.~Phys. A}{11}{1978}{983}
\bib{para}\ref{G Parisi}{Phys.~Lett.}{73}{1979}{203}
\bib{parb}\ref{G Parisi}{J.~Phys. A}{13}{1980}{L115}
\bib{parc}\ref{G Parisi}{J.~Phys. A}{13}{1980}{1101}
\bib{bisc}{\refs P Biscari}; in {\it The replicant model and the physics of
disordered systems,} Tesi di Perfezionamento at the Scuola Normale
Superiore, Pisa (1993).
\bye